\documentclass{article}
\usepackage{spconf,amsmath,graphicx,amssymb}
  \usepackage{subcaption}
  \usepackage{bm}
  \usepackage{multirow}
  \usepackage{booktabs}
  \usepackage{textgreek}
  \usepackage{gensymb}
  \usepackage{color,soul}

\title{Multi-resolution Location-based training for multi-channel continuous speech separation}
\name{Hassan Taherian$^1$,  {\normalfont{and}} DeLiang Wang$^{1,2}$ \thanks{This research was supported in part by two National Science Foundation grants (ECCS-1808932 and ECCS-2125074), the Ohio Supercomputer Center, and the Pittsburgh Supercomputer Center (NSF ACI-1928147).}}
\address{
$^1$Department of Computer Science and Engineering, The Ohio State University, USA\\
$^2$Center for Cognitive and Brain Sciences, The Ohio State University, USA\\	
\texttt{\small taherian.1@osu.edu, dwang@cse.ohio-state.edu}}
\begin{document}
\maketitle

\begin{abstract} 


The performance of automatic speech recognition (ASR) systems severely degrades when multi-talker speech overlap occurs. In meeting environments, speech separation is typically performed to improve the robustness of ASR systems. Recently, location-based training (LBT) was proposed as a new training criterion for multi-channel talker-independent speaker separation. Assuming fixed array geometry, LBT outperforms widely-used permutation-invariant training in fully overlapped utterances and matched reverberant conditions. This paper extends LBT to conversational multi-channel speaker separation. We introduce multi-resolution LBT to estimate the complex spectrograms from low to high time and frequency resolutions. With multi-resolution LBT, convolutional kernels are assigned consistently based on speaker locations in physical space. Evaluation results show that multi-resolution LBT consistently outperforms other competitive methods on the recorded LibriCSS corpus.

\end{abstract}

\section{Introduction}
	\label{sec:intro}

Multi-talker speaker separation methods based on deep neural networks (DNNs) have achieved impressive progress in recent years~\cite{wang2018supervised}. 
Existing separation methods are mainly concerned with fully overlapped utterances in short segments.
For practical applications, however, a separation model is required to process long audio recordings with occasional speech overlaps.
Recently, continuous speech separation (CSS) is proposed to handle overlapped speech in conversational or meeting environments~\cite{chen2020continuous}.
CSS usually divides a long audio stream into short segments. Each segment is independently processed by a separation model with permutation-invariant training (PIT) criterion. Separated signals are then concatenated to create individual speaker streams.

Many studies have been proposed to improve different aspects of the CSS framework~\cite{taherian2021time, Yixuan22, zhang22y_interspeech, neumann21_interspeech, Chenda, wang2021multi, saijo22b_interspeech, boeddeker22_interspeech}.
We introduced a modulation factor based on segment overlap ratio to dynamically adjust the separation loss~\cite{taherian2021time}. In \cite{Yixuan22}, a recurrent selective attention network is used to separate one speaker at a time. 
The work in \cite{zhang22y_interspeech} and \cite{neumann21_interspeech} proposed new training criteria that generalizes PIT to capture long speech contexts. 
Li et al.~\cite{Chenda} proposed a dual-path separation model to leverage inter-segment information from a memory embedding pool.
Wang et al. adopted a multi-stage approach by combining a multi-channel separation model with masking-based beamforming and a post-filtering network to achieve strong separation performance~\cite{wang2021multi}. 
Other works employed unsupervised multi-channel source separation for CSS.
Saijo et al. proposed a spatial loss that uses the estimated direction of arrival to impose spatial constraints on the demixing  matrix~\cite{saijo22b_interspeech}.
Boeddeker et al.~\cite{boeddeker22_interspeech} introduced an initialization scheme for a beamformer based on a complex Angular Central Gaussian Mixture model.

\begin{figure*}[!t]
	\centering
	\includegraphics[width=0.90\textwidth]{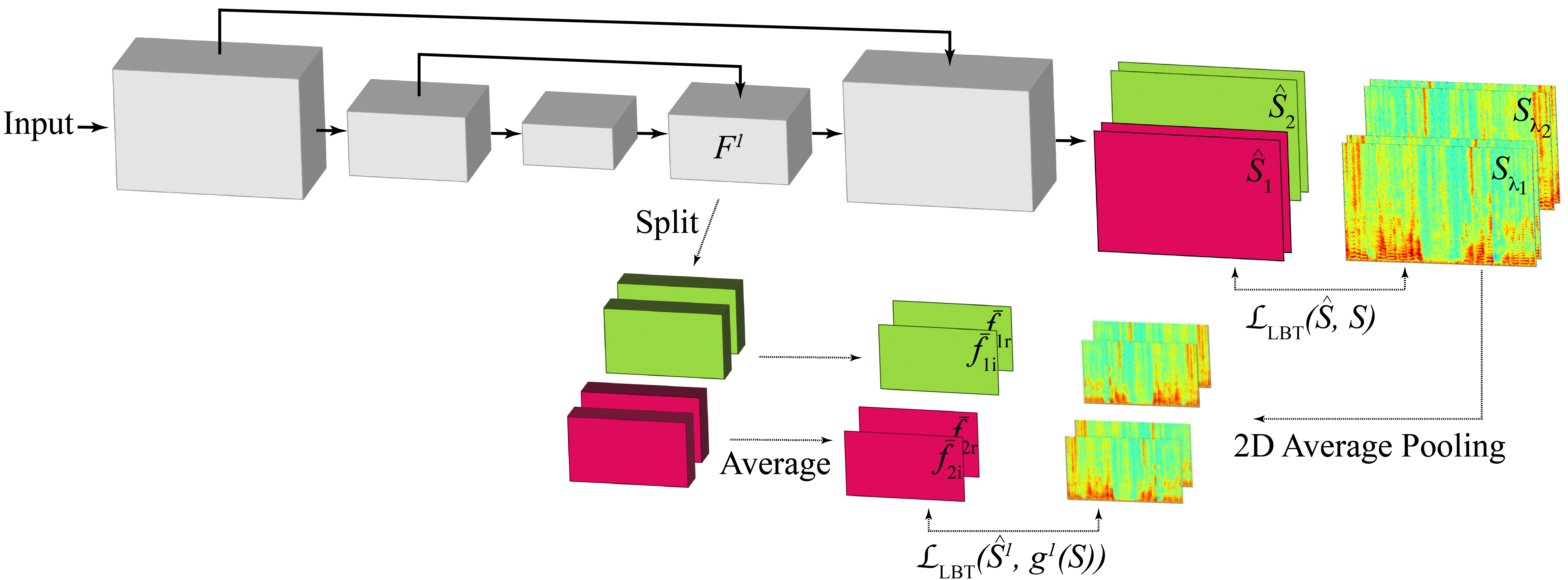}
	\caption{ \small
	  Schematic diagram of multi-resolution location-based training for a MC-CSM network with $K_e=3$ encoder layers and $K_d=2$ decoder layers.
	  A strided $2 \times 2$ depthwise convolutional layer is used after each 
	  dense block for downsampling and upsampling in encoder and decoder, respectively. Each cube represents an output feature map. Skip connections link encoders to the corresponding decoders. 
	  Feature maps in the decoder middle layers are divided into four groups and assigned to speakers according to their locations.}
	\label{fig_diagram}
	\end{figure*}
	

In our previous study, we introduced a new training criterion, named location-based training (LBT), to assign DNN outputs according to speaker locations in physical space~\cite{taherian2022Multi}. We showed that LBT performs better than PIT for fully overlapped utterances in simulated and matched reverberant conditions. This paper extends LBT to conversational multi-channel speaker separation and unmatched reverberant conditions. We train a multichannel input single-channel output (MISO) system which directly estimates the real and imaginary spectrograms of speakers from multi-channel mixture. We also propose a multi-resolution loss for convolutional neural networks by estimating the complex spectrograms from low to high time and frequency resolutions. With the multi-resolution loss, convolutional kernels are consistently paired with speakers according to their locations.
We demonstrate that multi-resolution LBT generalizes well to real conversational recordings, despite using simulated room impulse responses (RIRs) for training only. Experiments on the LibriCSS corpus~\cite{chen2020continuous} show that the proposed separation model produces superior separation performance compared to other competitive methods.

\section{System Description}
	\label{sec:sys_description}
\subsection{Location-based training for CSS}
	\label{sec:location_based_training}

The objective of CSS is to generate multiple non-overlapped streams from a multi-channel signal that contains occasional speech overlaps~\cite{chen2020continuous}. The CSS framework can handle any number of speakers as long as there are at most $N$ active speakers in a segment at the same time. In this study, we assume $N=2$ as more than two concurrent speakers rarely happens for a short length segment. Most prior works in CSS employ utterance-wise PIT for training the separation model~\cite{chen2020continuous, wang2021multi, Chen2021conformer}:
\begin{equation}
	\mathcal{L}_{\text{PIT}}(\hat{S}, S)  = \min_{\phi \in \Phi}  \sum_{n=1}^{N} \mathcal{L}(\hat{S}_n, S_{\phi(n)}),
	\label{eq_pit_loss}
 \end{equation}
where $\hat{S}$ and $S$ are estimated and clean speech signals in the short-time Fourier transform (STFT) domain, respectively.
$\mathcal{L}$ denotes a loss function, symbol $\Phi$ the set of all permutations and $\phi$ refers to one permutation. For the segments with no speech overlap, the PIT-based separation model puts the input to one of its outputs while the other output channel produces a zero signal.

Recently LBT was proposed to resolve the permutation ambiguity problem in multi-channel talker-independent speaker separation~\cite{taherian2022Multi}. LBT leverages distinct spatial locations of multiple speakers in physical space and produces superior separation performance compared to PIT. Moreover, LBT has a linear training complexity to the number of speakers which computationally much more efficient than PIT with a factorial complexity. In this study, we extend the LBT criterion to CSS:

\begin{equation}
	\mathcal{L}_{\text{LBT}}(\hat{S}, S)  = \sum_{n=1}^{N} \mathcal{L}(\hat{S}_n, S_{\lambda_{n}}),
	  \label{eq_azi_loss}
\end{equation}
where $ \lambda_{n} \in \{1, 2\}$  is speaker index sorted in ascending order based on speaker azimuths or distances relative to the microphone array. Different from PIT, Eq.~\eqref{eq_azi_loss} organizes DNN outputs on the basis of speaker azimuths or distances. For non-overlap segments, the input is mapped to the first output and a zero signal is assigned to the second output.

\subsection{Multi-channel complex spectral mapping}

Our separation model is based on multi-channel complex spectral mapping (MC-CSM) which achieves better separation performance than masking-based beamforming~\cite{williamson2016complex, neuralSpatialFilter}. We employ the Dense-UNet architecture proposed in \cite{liu2019divide} for MC-CSM. The Dense-UNet input is the real and imaginary components of the mixture STFT at all microphones as well as the spectral magnitude of the first microphone. The Dense-UNet contains an encoder and decoder with $K_e=5$ and $K_d=4$ layers of densely-connected convolutional blocks, respectively.
Each dense block has 5 convolutional layers with 76 channels, a kernel size of $3 \times 3$, and a stride of $1 \times  1$. After the last dense block, we use a $1 \times  1$ convolutional layer with 4 channels to produce estimates of the real and imaginary components for each speaker. MC-CSM contains around 6.9 million parameters.
For more details about the Dense-UNet architecture, we refer the reader to \cite{liu2019divide}. The model is trained with the loss function proposed by~\cite{wang2020multijournal}:
\begin{equation}
  \begin{split}
	\mathcal{L}(\hat{S}, S) &=  \left\lVert \hat{S}^{(r)} - S^{(r)} \right\rVert_{1} + \left\lVert \hat{S}^{(i)} - S^{(i)} \right\rVert_{1} \\
   &+ \left\lVert |\hat{S}| - |S| \right\rVert_{1},
  \end{split}
\end{equation}
where superscripts $r$ and $i$ denote real and imaginary parts, $|~.~|$ computes magnitude and $ \left\lVert~.~\right\rVert_{1}$ computes $\ell_1$ norm.

\subsection{Multi-resolution location-based training}

The key idea of multi-resolution LBT is to estimate the complex 
spectrograms using dedicated convolutional kernels in the decoder by progressively increasing the time and frequency resolutions. Fig.~\ref{fig_diagram} illustrates  multi-resolution LBT for a Dense-UNet  with $K_e=3$ encoder layers and $K_d=2$ decoder layers. For every middle layer $k$ in the decoder, we split the channel dimension of the output feature maps $F^k$ into four groups:
\begin{equation}
f_{1r}^k, f_{1i}^k, f_{2r}^k, f_{2i}^k  = \text{Split}(F^k)
	  \label{eq_split}
\end{equation}
where $f_{nr}^k$ and $f_{ni}^k$ are the real and imaginary feature maps for speaker $n\in \{1, 2\}$, respectively. Then, the feature maps within each group are averaged across the channel dimension.
A low-resolution estimate of the speech signals in the STFT domain is created at layer $k$ for each speaker:
\begin{equation}
	\hat{S}_n^k  = \bar{f}_{nr}^k +i \bar{f}_{ni}^k
	  \label{eq_average}
\end{equation}
where $\bar{f}$ is the averaged feature maps and $i$ denotes the imaginary unit. The LBT loss between lower resolution estimates and clean signals is calculated for every middle layer and added to the final loss: 

\begin{equation}
	\mathcal{L}_{\text{LBT}}(\hat{S}, S) + \sum_{k}^{K_d-1} \mathcal{L}_{\text{LBT}}(\hat{S}^k, g^{K_d-k}(S))  
	  \label{eq_avg_pool}
\end{equation}

\noindent where $g^x(.)$ is a 2D average pooling function with a kernel size of $2 \times 2$ and a stride of $2 \times  2$, applied recursively for $x$ iterations. 
Intuitively, we assign a group of convolutional kernels to learn the real and imaginary spectrograms of each speaker with different time and frequency resolutions. The kernel assignment is consistent at every decoder layer based on sorted speaker azimuth angles or distances.

Note that this is different from multi-scale loss~\cite{nachmani2020voice} which enforces producing a waveform estimate after each layer by using a shared decoder for a PIT-based separation model. We emphasize that consistent assignments of convolutions kernels is essential for the multi-resolution loss. We expect that combining the multi-resolution loss with PIT criterion would not perform well as PIT allows label permutation and the order of convolutions kernels can change from one mixture to another.

\begin{table*}   
	\caption{WER results (in \%) of comparison systems for utterance-wise and continuous evaluation on LibriCSS. `MISO\textsubscript{1}+SC' and `MC-CSM+SC' refer to MISO\textsubscript{1} and MC-CSM with a speaker counting network, respectively. }

	\label{tab:res}
	\centering
	\renewcommand{\arraystretch}{1.00}
	\resizebox{0.95\textwidth}{!}{
	  \begin{tabular}{{l c c cccccc | cccccc  }}
	  \toprule
	  \multicolumn{3}{l}{\multirow{2}{*}{}}  & \multicolumn{6}{c}{\textbf{Utterance-wise}} & \multicolumn{6}{c}{\textbf{Continuous}}  \\
	  \cmidrule{4-9} \cmidrule{10-15}
	  & Criterion & \shortstack{Multi- \\ resolution}   & 0S & 0L & 10\%  & 20\% & 30\% & 40\%   & 0S & 0L & 10\%  & 20\% & 30\% & 40\%  \\  
	  \midrule
	  Unprocessed 									&--	&--				& 11.8		&11.7	&18.8	&27.2	&35.6	&43.3 
																	  &15.4		&11.5	&21.7	&27.0		&34.3	&40.5\\
																
	  BLSTM~\cite{chen2020continuous}    		&PIT&--			& 8.3		&8.4	&11.6	&16.0		&18.4	&21.6
																	  &11.9		&9.7	&13.4	&15.1	&19.7	&22.0\\

	  Conformer~\cite{Chen2021conformer}    		&PIT&--			& 7.2		&7.5	&9.6	&11.3	&13.7	&15.1	
																	&11.0			&8.7	&12.6	&13.5	&17.6	&19.6\\

	  MISO\textsubscript{1}~\cite{wang2021multi}	&PIT&--			& 7.7		&7.5	&7.9	&9.6	&11.3	&13.0	
																	  &10.7		&10.5	&10.9	&11.5	&13.8	&15.3\\

	  MISO\textsubscript{1}+SC~\cite{wang2021multi}	&PIT&--			& --		&--		&--		&--		&--		&--	
																	  &7.9		&8.5	&8.5	&10.5	&12.3	&14.3\\
	  \midrule 
	  \multirow{5}{*}{MC-CSM}						&PIT&-- 			&7.0	&7.5	&7.9	&9.3	&11.3	&13.0
																	  &10.0			&9.4	&11.1	&11.8	&14.3	&15.2 \\

	  												&Azimuth&--		&8.6		&8.8	&9.2	&10.3	&11.3	&12.4
																	  &13.5		&14.5	&13.6	&13.4	&14.7	&15.4 \\

	  		 										&PIT&\checkmark		&7.6	&7.6	&7.8	&9.4	&11.0	&12.8
																	  &11.4	    &12.5	&11.3	&13.0	&15.1	&16.3 \\
																	  
	  		 										&Azimuth&\checkmark		&7.3		&7.9	&7.6	&\textbf{9.2}	&\textbf{10.8}	&\textbf{11.4}
																	  &9.2		&9.8	&9.1	&10.5	&12.2	&\textbf{12.8} \\

	  		 										&Distance&\checkmark	&8.7		&10		&10.5	&12.9	&14.6	&17.8	
												   					&12.2		&12.8	&12.4	&14.8	&17.1	&19.0	 \\
	  \midrule 

	\multirow{5}{*}{MC-CSM+SC}						&PIT &--			&\textbf{6.5}		&\textbf{6.9}	&7.4	&9.3	&11.7	&13.6
																	  &8.0			&\textbf{7.4}	&8.8	&10.7	&12.6	&13.8 \\

	  												&Azimuth&--			&7.4		&7.7	&8.1	&9.5	&11.3	&12.5 
																		  &7.9		&7.9	&8.5	&9.4	&\textbf{11.6}	&12.9\\

	  		 										&PIT&\checkmark		&\textbf{6.5}	    &7.3	&\textbf{7.3}	&9.6	&11.3	&13.5
																	    &8.1	    &8.1	&8.3	&10.2	&12.4	&14.0 \\
																	  
	  			 									&Azimuth&\checkmark			&6.7		&7.1	&7.5	&\textbf{9.2}	&10.9	&12.1
																		  &\textbf{7.7}		&7.7	&\textbf{8.0}		&\textbf{9.3}	&\textbf{11.6}	&13.0 \\

													&Distance&\checkmark		&7.6	&8.3	&9.3	&12.1	&14.2	&17.8	
																		  &8.3	&8.4	&9.4	&11.7	&14.9	&16.8	 \\
	  \bottomrule
  
	  \end{tabular}
	}
  \end{table*}

\section{Experimental Setup}
	\label{sec:exp_setup}

\subsection{Datasets}\label{sec_dataset}

We evaluate the proposed separation models with the LibriCSS corpus~\cite{chen2020continuous} which is a conversational speech recognition task with speech overlaps. This dataset contains 10 hours of recordings from the Librispeech development set that are retransmitted with loudspeakers to capture real room reverberation. 
The loudspeakers are positioned around a table and their distance to the array ranges from 33 to 409 cm.
The recordings are divided into 6 sessions with different overlap ratios: 0S (no overlap with a 0.1-0.5s pause between utterances), 0L (no overlap with a 2.9-3.0s pause between utterances), 10\%, 20\%, 30\% and 40\% overlaps. The LibriCSS corpus is recorded using a 7-channel circular microphone array with 4.25 cm radius.

We follow the setup described in \cite{wang2021multi} for simulating training and validation data using the image method~\cite{allen1979image, scheibler2018pyroomacoustics}. We created 192K two-speaker mixtures with different overlap ratios using the LibriSpeech dataset. Simulating the LibriCSS recording device, each mixture is convolved with 7-channel microphone array RIRs. For generating RIRs, we simulate rectangular rooms with random length, width and height dimensions in the range of [5$\times$5$\times$3, 10$\times$10$\times$4] meters, with the microphone array placed in the center of the room. The speaker positions are uniformly sampled from 360 candidate azimuth angles in the range of -180\degree~ to 180\degree~ with a 1\degree~ resolution. The speaker-array distances are randomly selected such that the two speakers are at least 5 cm far apart. The reverberation time (T60) is randomly sampled between 0.2 and 0.6 s. Stationary ambient noise is also added to the mixture signal at a random SNR from 10 to 30 dB. 

The separation models are evaluated using the default ASR system provided with the LibriCSS corpus~\cite{chen2020continuous}. The LibriCSS contains two evaluation scenarios: utterance-wise and continuous evaluations. In the utterance-wise scenario, word error rate (WER) is calculated for both separated signals of pre-segmented mixtures, and the one with lower WER is considered. The continuous scenario evaluates long mixture segments that contain 8-10 utterances. The decoding results from both streams are combined for evaluation.

For both scenarios, the input mixture is processed in 2.4 s segments with a segment shift of 1.2 s. 
The processed segments are concatenated using the stitching algorithm proposed by \cite{chen2020continuous}. All signals are sampled at 16 kHz. We use STFT with a frame length of 32 ms and a frame shift of 8 ms. The sample variance of the multi-channel input mixture is normalized. Global mean and variance normalization is also applied to STFT features.

\subsection{Speaker counting}\label{sec_source_counting}
For the segments without speaker overlaps, the separation model is expected to produce a zero signal in one of its outputs. However, it is observed that the model sometimes emits an intelligible residual signal for non-overlapped segments, leading to more decoding errors. Following~\cite{wang2021multi}, we utilize a frame-level speaker counting (SC) to find non-overlapped frames and remove the signal residual by combining the two separated outputs. For each frame, SC produces the probabilities for three classes: zero, one or two speakers. The SC model is based on a modified  MC-CSM architecture. Specifically, SC consists of the MC-CSM encoder followed by two bi-directional long short-term memory (BLSTM) layers and a softmax layer. The SC labels for training were generated using a pre-trained DNN based voice activity detector~\cite{Kaldi_VAD}.

\section{Evaluation Results}\label{sec:eval}

We trained MC-CSM using PIT and LBT criteria with and without the multi-resolution loss. For LBT, output-speaker assignments are ordered based on speaker azimuth angles. Table~\ref{tab:res} presents the WER results of MC-CSM models for utterance-wise and continuous evaluations. 
We observe MC-CSM trained with azimuth criteria yields comparable results to PIT in higher overlap ratios but it performs worse than PIT in lower overlap ratios.
However, clear improvement is observed with azimuth-based training when SC is used to suppress the residual signal in non-overlapped segments. Combining the multi-resolution loss and azimuth-based training shows excellent speech recognition performance and outperforms PIT with or without speaker counting.
For example, MC-CSM based on multi-resolution LBT leads to a 2.4\% absolute WER reduction compared to PIT on the 40\% overlap condition in continuous evaluation.
The results indicate that LBT models trained with simulated RIRs generalize well to real microphone array recordings. 
On the other hand, PIT-based separation model performance does not improve when the multi-resolution loss is used. This is because the PIT criterion is order-agnostic and features maps are not sorted consistently with the multi-resolution loss for different mixtures.

We also trained MC-CSM with LBT by using speaker distances instead of azimuth angles. As shown in the table, distance-based training underperforms both azimuth-based training and PIT. The reason could be that the distance criterion is not discriminative enough in the LibriCSS dataset as loudspeakers are positioned symmetrically and their distances to the microphone array are close.

Finally, we compare our MC-CSM models with other competitive separation methods that are evaluated with the LibriCSS default ASR. For all methods, we list the best reported results, and leave unreported fields blank.  The proposed systems in \cite{chen2020continuous} and \cite{Chen2021conformer} respectively use BLSTM and conformer architectures to estimate real-valued masks. The system in \cite{wang2021multi} performs complex spectral mapping using a UNet model with a temporal convolutional network. A dedicated speaker counting network was also utilized to further improve WER scores. Our LBT models trained with the  multi-resolution loss produce much better WER scores compared to other systems. We should mention that WER scores can be further decreased by using a speech enhancement network on top of the separation model as done in~\cite{wang2021multi}. Our focus in this study is on the separation model, and further improvements can be expected by introducing speech enhancement. 

\section{Concluding Remarks}
	\label{sec:conclude}

In this study, we have extended LBT to conversational multi-channel speaker separation. We have proposed a multi-resolution loss to consistently assign convolutional kernels according to speaker locations. We have demonstrated LBT yields significantly better separation and ASR performance than other competitive methods on the LibriCSS corpus.



\bibliographystyle{IEEEtran}
{\small\bibliography{mybib}}
\end{document}